\documentclass[journal]{IEEEtran} 

\ifCLASSINFOpdf
\else
\fi

\usepackage{color}
\usepackage{amssymb}
\usepackage{amsthm}
\usepackage[cmex10]{amsmath}
\DeclareMathOperator{\E}{\mathbb{E}}

\DeclareMathOperator{\C}{\mathbb{C}}

\newcommand {\Define} {\stackrel {\Delta} {=}  }

\newcommand{\mya}{\mathrel{\overset{\makebox[0pt]{{\tiny(a)}}}{=}}}

\newcommand {\pu} {p_{\text{u}}}

\newcommand {\gamab} {\gamma_{\text{b}}}
\newcommand {\gamabp} {\gamma_{\text{b,pcsi}}}
\newcommand {\gamabi} {\gamma_{\text{b}}(R,R^{\prime})}

\newcommand {\Rpcsi} {R_{\text{pcsi}}}
\newcommand {\Rcsi} {R_{\text{icsi}}}
\newcommand {\gamapcsi} {\gamma_{\text{pcsi}}^{0}}
\newcommand {\gamacsi} {\gamma_{\text{icsi}}^{0}}
\newcommand {\Rpcsit} {\widetilde{R}_{\text{pcsi}}}

\usepackage{bm}
\usepackage{mathtools}
\usepackage{fixltx2e}
\usepackage{epsfig,makeidx,color}
\usepackage{graphicx,dblfloatfix}
\usepackage{amsbsy}
\usepackage{amssymb}
\usepackage{euscript}
\usepackage{lipsum}
\usepackage{cite}
\usepackage{chngcntr}
\usepackage{lineno}
\usepackage[T1]{fontenc}
\usepackage{microtype}
\usepackage{wasysym}
\usepackage{footnote}
\newtheorem{theorem}{Theorem}

\newtheorem{proposition}{Proposition}

\newtheorem{remark}{\it Remark}

\usepackage[english]{babel}
    \usepackage[font=small,labelsep=space]{caption}
\usepackage{subcaption}    
\makeatletter
\def\citenoauxwrite#1{\begingroup
\@fileswfalse
\cite{#1}\relax
\endgroup}
\makeatother

\begin{document}

\title{How Much Bandpass Filtering is Required in Massive MIMO Basestations?}
%
%
%

\author{Sudarshan~Mukherjee~and 
        ~Saif Khan~Mohammed
\thanks{The authors are with the Department of Electrical Engineering, Indian Institute of Technology (I.I.T.) Delhi, India. Saif Khan Mohammed is also associated with Bharti School of Telecommunication Technology and Management (BSTTM), I.I.T. Delhi. Email: saifkmohammed@gmail.com. This work is supported by EMR funding from the Science and Engineering
Research Board (SERB), Department of Science and Technology (DST),
Government of India.}
}

\maketitle


\begin{abstract}
In this paper, we study the impact of aliased out-of-band interference signals on the information sum-rate of the maximum ratio combining receiver in massive multiple-input multiple-output (MIMO) uplink, with both perfect and imperfect channel estimates, in order to determine the required out-of-band attenuation in RF bandpass filters (BPFs). With imperfect channel estimates, our study reveals that as the number of base-station (BS) antennas ($M$) increases, the required attenuation at the BPFs increases as $\mathcal{O}(\sqrt{M})$ with $M \to \infty$, provided the desired information sum-rate (both in the presence and in the absence of AOOBIs (aliased out-of-band interferers)) remains fixed. This implies a practical limit on the number of BS antennas due to the increase in BPF design complexity and power consumption with increasing $M$.
\end{abstract}



%

\vspace{-0.4 cm}

\section{Introduction}
%
%
%
%
In the development of the next generation ($5$G) wireless communication systems, massive multiple-input multiple-output (MIMO) has been visualized as a key technology, which would supplement other $5$G technologies forming an integrated communication network, which has high energy and spectral efficiency and low latency \cite{Andrews}. The vision for massive MIMO is to equip the base-station (BS) with a large antenna array (of the order of hundreds) to support a few tens of users in the same time-frequency resource \cite{Marzetta1}. For massive MIMO systems, it has been suggested that low-complexity signal processing can achieve good information rate performance due to the averaging of noise and hardware imperfections across the $M$ BS antennas (as $M \to \infty$) \cite{Marzetta2,Emil6}.

\par
\begin{figure}[t]
\vspace{-0.4 cm}
\centering
\includegraphics[width= 3.4 in, height= 1.2 in]{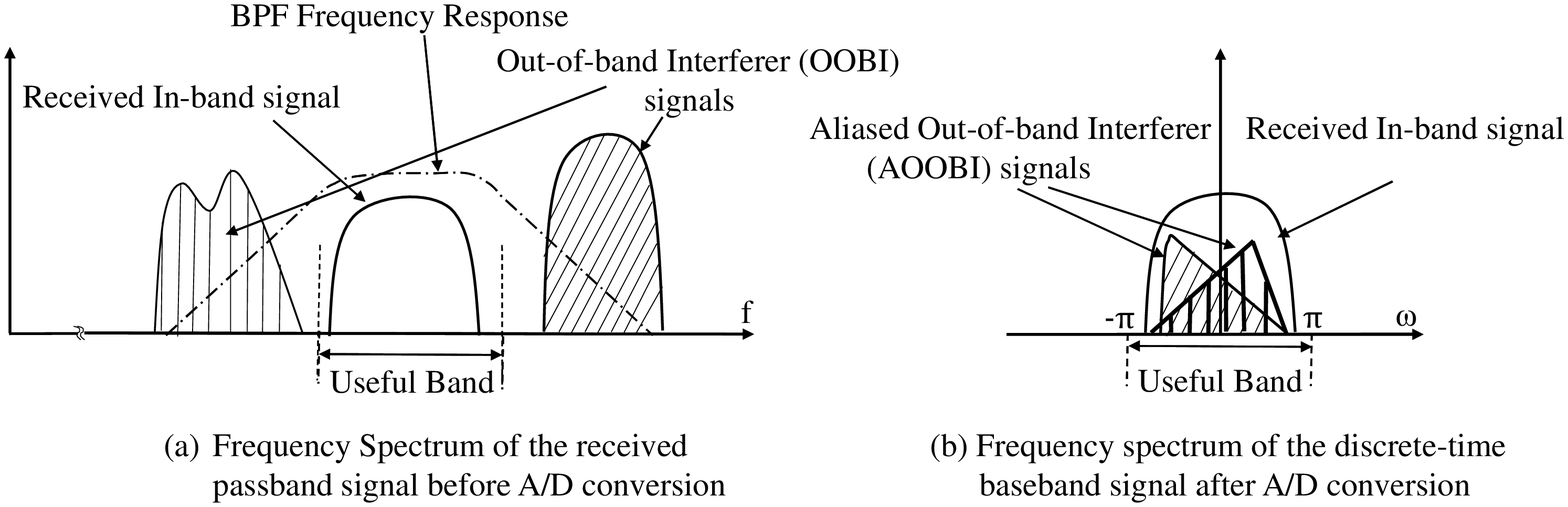}
\caption {Aliasing of OOBI Signals due to insufficient filtering.} 
\label{fig:motivate}
\vspace{-0.6 cm}
\end{figure}

 All these results for massive MIMO system, however, have been studied under the assumption of ideal bandpass filtering in the RF chain of each BS antenna. In a communication receiver, the received passband signal is first filtered through a bandpass filter (BPF) to remove out-of-band interferers (OOBIs) and noise. In the design of BPFs, sufficient attenuation is required in the out-of-band (OOB) regions compared to the communication band of interest (also referred to as the useful band/in-band). Insufficient OOB attenuation would result in aliasing of the filtered OOBI signals into the useful band, at the output of the analog-to-digital converter (ADC) (see Fig.~\ref{fig:motivate}). Presence of such \textit{aliased OOBI} (AOOBI) signals in the useful band would corrupt the received baseband signal of interest, thereby degrading the system performance. In this paper we address the question of exactly how much BPF attenuation is required in the OOB regions, or equivalently what is the maximum allowable ratio (MAR) of the total AOOBI power in the useful band to the average received in-band power before aliasing (RIBP) in order to achieve a given fractional loss in the information sum-rate.\footnote[1]{The fractional loss in the information sum-rate is defined as the ratio of the loss in the sum-rate in the presence of AOOBIs to the sum-rate in their absence.} {To the best of our knowledge, this is the first paper to report such a study in massive MIMO systems}.


\textsc{Contributions}: The novel contributions of our work in this paper are as follows: (i) we have derived closed-form expressions for the information sum-rate of the maximum ratio combining (MRC) receiver in massive MIMO uplink in the presence of AOOBIs, for both the perfect and imperfect CSI (channel state information) scenarios; (ii) For the imperfect CSI scenario, our analysis of the information sum-rate expression reveals that for a fixed desired information sum-rate in the absence of AOOBIs and a fixed desired fractional loss due to the presence of AOOBIs, the MAR decreases as $\frac{1}{\sqrt{M}}$ with $M \to \infty$ (see Theorem~\ref{imcsifiltcut} in Section III-B). This is in contrast to the result in the perfect CSI scenario, where the MAR converges to a constant as $M \to \infty$ (see Proposition~\ref{pcsifiltcut} in Section II-B). {The $\mathcal{O}(\sqrt{M})$ decrease in the MAR, or equivalently the $\mathcal{O}(\sqrt{M})$ increase in the required BPF attenuation in the OOB region results in increased hardware design complexity, cost and power consumption. This therefore limits the number of BS antennas that can be employed in practice.\footnote[2]{{Possible future work may include extension of this work to multi-cell scenario.}}} [\textbf{{Notations:}} $\C$ denotes the set of complex numbers. $\E$ denotes the expectation operator. $(.)^{H}$ denotes the complex conjugate transpose operation, while $(.)^{\ast}$ denotes the complex conjugate operator and $(.)^T$ denotes transpose operation.] 

\vspace{-0.4 cm}

\section{System Model}

Let us consider a single-cell single-carrier massive MIMO cellular BS with $M$ antennas, serving $K$ single antenna user terminals (UTs) in the same time-frequency resource. We also assume that duration of the coherence interval is $N_c$ channel uses, of which $N_u$ channel uses are allocated for the uplink (UL) transmission. The independent channel gain coefficient from the single antenna of the $k^{\text{th}}$ UT to the $m^{\text{th}}$ BS antenna is $h_{mk} \sim \mathcal{C}\mathcal{N}(0, \beta_k)$, where $\sqrt{\beta_k}>0$ is the geometric attenuation coefficient ($m = 1, 2, \ldots, M$ and $k = 1, 2, \ldots, K$). Let the transmitted symbol from the $k^{\text{th}}$ UT at any time instance $t$ in the UL be denoted by $s_k[t]$ ($t = 0, 1, 2, \ldots, N_u - 1$), and let the average power transmitted from each UT be $\E[\, |s_k[t]|^2 \,] = \pu$. {We model single antenna out-of-band interferers (OOBIs), with the independent and identically distributed (i.i.d.) channel gain coefficient from the $i^{\text{th}}$ OOBI to the $m^{\text{th}}$ BS antenna denoted by $g_{i,m} \sim \mathcal{C}\mathcal{N}(0,1)$, where $i = 1, 2, \ldots, \mathcal{I}$ ($\mathcal{I}$ is the total number of OOBIs). Also $g_{i,m}$ are independent of $h_{mk}$, for all $m = 1, 2, \ldots, M$ and $k = 1, 2, \ldots, K$. Further let $u_i[t] \sim \mathcal{C}\mathcal{N}(0,p_i)$ denote the independent Gaussian symbols (independent across $i$ and $t$; $i = 1, 2, \ldots, \mathcal{I}$) transmitted by the $i^{\text{th}}$ OOBI at time $t$ and $p_i$ is the average in-band power received at each BS antenna due to aliasing of the $i^{\text{th}}$ interferer into the useful band of interest.} Let $r_m[t]$ denote the signal received at the $m^{\text{th}}$ BS antenna at time $t$. Then $r_m[t]$ is given by

\vspace{-0.6 cm}

{\begin{IEEEeqnarray}{rCl}
\label{eq:rxsig}
r_m[t]  =  \sum_{q = 1}^{K} h_{mq} s_q[t] \, + \, \sum_{i = 1}^{\mathcal{I}} g_{i, m}u_i[t] \, + \, w_m[t],
\end{IEEEeqnarray}}

\vspace{-0.3 cm}

\noindent where $t = 0, 1, \ldots, N_u - 1$ and $m = 1, 2, \ldots, M$. Here $w_m[t] \sim \mathcal{C}\mathcal{N}(0, \sigma^2)$ is the i.i.d. circular symmetric additive white Gaussian noise (AWGN) at the $m^{\text{th}}$ BS antenna.

\vspace{-0.5 cm}

\subsection{Measure of the Required OOB Attenuation}

Before aliasing of the OOBIs, the total received in-band power at each BS antenna is given by $P_I \Define \E\big[ \, \big|\sum_{q = 1}^{K} h_{mq}s_q[t] + w_m[t] \big|^2 \, \big] = \big(\pu \sum_{q = 1}^{K} \beta_q + \sigma^2\big)$. Let the power of the $i^{\text{th}}$ OOBI received at each BS antenna before bandpass filtering be denoted by $ p_i^{\prime}$. We assume an ideal bandpass filter (BPF) which has a unit gain in the useful band and a gain $A<1$ outside the useful band of interest (i.e. OOB region). Then at the BPF output, the power received from the $i^{\text{th}}$ OOBI would be reduced to $p_i \Define A p_i^{\prime}$. This power would then be aliased into the useful band at the ADC output (i.e. during discrete time sampling). Ideally we would like to choose $A <1$ to be as large as possible so as to reduce the complexity of the BPF. Therefore in this paper, we seek to find the largest possible $A$ (i.e. the smallest required BPF attenuation in the OOB region) for a given allowable fractional loss in the information sum-rate of the massive MIMO system. This is equivalent to finding the maximum allowable ratio (MAR) of the total aliased OOBI (AOOBI) power to the total received in-band power before aliasing (RIBP), i.e.,

\vspace{-0.55 cm}

{\begin{IEEEeqnarray}{rCl}
\label{eq:filtcut}
\frac{ \sum_{i = 1}^{\mathcal{I}}A \, p_{\text{i}}^{\prime}}{\pu \sum_{q = 1}^{K}\beta_q + \sigma^2} = \frac{ \gamma_{\text{b}}}{\gamma \sum_{q = 1}^{K}\beta_q + 1} \Define r_{\text{b}},
\IEEEeqnarraynumspace
\end{IEEEeqnarray}}

\vspace{-0.2 cm}

\noindent for a given fractional loss in the information sum-rate. {Here $\gamma \Define \frac{\pu}{\sigma^2}$, $\gamma_{\text{b}} \Define \sum_{i = 1}^{\mathcal{I}}\gamma_i$ and $\gamma_i \Define \frac{p_i}{\sigma^2} = \frac{ A \, p_i^{\prime}}{\sigma^2}$. The ratio $r_{\text{b}}$ is a measure of the required BPF attenuation in the OOB region.\footnote[3]{{Let us consider an LTE like system with a channel bandwidth 1.4 MHz. The 3GPP LTE specifications for adjacent channel selectivity (ACS) conformance rule\cite{3GPP2} specifies a total $31.5$ dB more interference power than the power of the received useful signal. If the MAR for a $5\%$ fractional loss in sum-rate is $-15$ dB, then it is clear that the BPF must provide an attenuation of at least $31.5 + 15 = 46.5$ dB in the adjacent band.}}}

\subsection{MAR Analysis in Perfect CSI Scenario}

With perfect CSI (i.e. BS has perfect knowledge of $h_{mq}$, $q = 1, 2, \ldots, K$), we have $s_k[t] = \sqrt{\pu} \, x_k[t], \, \forall t = 0, 1, \ldots, N_u - 1$ and $k = 1, 2, \ldots, K$, where $x_k[t]$ is the i.i.d. Gaussian information symbol transmitted from the $k^{\text{th}}$ UT at time $t$, i.e., $x_k[t] \sim \mathcal{C}\mathcal{N}(0,1)$. For the $k^{\text{th}}$ UT, the output of the MRC receiver at the BS is given by

\vspace{-0.6 cm}

{\begin{IEEEeqnarray}{rCl}
\label{eq:detsigpcsi}
\nonumber \widehat{x}_k[t] & = & \sum\limits_{m = 1}^{M}h_{mk}^{\ast}r_m[t] \, \mya \, \sqrt{\pu}||\, \bm h_k \,||^2 \, x_k[t] \,+\, \underbrace{\bm h_k^H \bm w[t]}_{\Define \, \text{EN}_k[t]}\\
\nonumber & & \,\,\,\,\,\,\,\,\,\,\,\,\,\,\,\,\,\, + \underbrace{\sqrt{\pu}\sum\limits_{q = 1, q\neq k}^{K}\bm h_k^H \bm h_q x_q[t]}_{\Define \, \text{MUI}_k[t]} \,+\, \underbrace{\sum\limits_{i = 1}^{\mathcal{I}} \bm h_k^H \bm  g_i u_i [t]}_{\Define \, \text{BL}_k[t]} \\
\nonumber & = & \underbrace{\sqrt{\pu}\E\Big[||\bm h_k||^2\Big] \, x_k[t]}_{\Define \, \text{ES}_k[t]} + \underbrace{\sqrt{\pu} \,(||\bm h_k||^2 - \E[\,||\bm h_k||^2\,]\,)x_k[t]}_{\, \Define \, \text{SIF}_k[t]}\\
& & \,\,\,\,\,\,\,\,\,\,\,\,\,\,\,\,\, + \text{MUI}_k[t] + \text{BL}_k[t] + \text{EN}_k[t],
\IEEEeqnarraynumspace
\end{IEEEeqnarray}}

\vspace{-0.5 cm}

\noindent  where step $(a)$ follows from \eqref{eq:rxsig}. Here $\text{SIF}_k[t]$ is the self-interfering signal component and $\text{EW}_k[t] \Define \text{SIF}_k[t] + \text{MUI}_k[t] + \text{BL}_k[t] + \text{EN}_k[t]$ is the overall interference and noise term. Here, $\bm h_q \Define (h_{1q}, h_{2q}, \cdots, h_{Mq})^T$, $q = 1, 2, \ldots, K$. Also, $\bm w[t] \Define (w_1[t], w_2[t], \cdots, w_M[t])^T$ and $\bm g_i \Define (g_{i,1}, g_{i,2}, \cdots, g_{i,M})^T$. Note that $\E[\text{EW}_k^{\ast}[t]x_k[t]] = 0$, $k = 1, 2, \ldots, K$ and $t = 0, 1, \ldots, N_u - 1$, i.e., the effective noise $\text{EW}_k[t]$ is uncorrelated with the Gaussian information symbol $x_k[t]$. With Gaussian information symbols, the worst case uncorrelated noise (in terms of mutual information) is also Gaussian \cite{Hasibi2} with the same variance as that of $\text{EW}_k[t]$. Therefore, $I(\widehat{x}_k[t]; x_k[t]) \geq \log_2 (1 + \text{SINR}_{k, \text{pcsi}})$, where 

\vspace{-0.45 cm}

\begin{IEEEeqnarray}{rCl}
\label{eq:sinrpcsi}
\text{SINR}_{k,\text{pcsi}} & \Define & \frac{\E[\,|\text{ES}_k[t]|^2\,]}{\E[\,|\text{EW}_k[t]|^2\,]} \, = \, \frac{M}{1 + \frac{1 + \gamma_{\text{b}}}{\beta_k \gamma} + \sum_{q = 1, q\neq k}^{K} \frac{\beta_q}{\beta_k}}.
\IEEEeqnarraynumspace
\end{IEEEeqnarray}

\vspace{-0.15 cm}

\indent Here the expectation is over the fast fading component of channel gains and AWGN and {$\gamma_{\text{b}} = \sum_{i = 1}^{\mathcal{I}} \gamma_i$. Using \eqref{eq:sinrpcsi}, the information sum-rate in the presence of aliased OOBIs (AOOBIs) is a function of $\gamma = \frac{\pu}{\sigma^2}$, $\bm \Gamma_{\text{b}} \Define (\gamma_1, \gamma_2, \cdots, \gamma_{\mathcal{I}})^T$ and is given by}

\vspace{-0.6 cm}

{{\begin{IEEEeqnarray}{rCl}
\label{eq:pcsisumrate}
\Rpcsi (\gamma, \bm \Gamma_{\text{b}})  \Define  \sum\limits_{k = 1}^{K} \log_2 \Bigg(1 + \frac{1}{\frac{\big(1 + \sum_{q = 1, q\neq k}^{K}\frac{\beta_q}{\beta_k}\big)}{M } + \frac{1 + \bm 1^T \bm \Gamma_{\text{b}}}{\beta_k M \, \gamma}}\Bigg).
\IEEEeqnarraynumspace
\end{IEEEeqnarray}}}

\vspace{-0.3 cm}

\indent {Note that\footnote[4]{{$\bm 1$ and $\bm 0$ denote the all $1$ and all $0$ column vectors respectively.}} $\Rpcsi$ depends on $\bm \Gamma_{\text{b}}$ only through $\bm 1^T \bm \Gamma_{\text{b}} \Define \gamab$. Clearly, for any $0 < R < \Rpcsi(\infty, \bm 0)$ [where $\Rpcsi(\infty, \bm 0) \Define \lim\limits_{\gamma \to \infty} \, \Rpcsi(\gamma, \bm \Gamma_{\text{b}} = \bm 0)$], we define $\gamma = \gamapcsi(R)$ to be the unique solution to the equation $\Rpcsi(\gamma,\bm 0) = R$. Therefore for any sum-rate $R$, $\gamapcsi(R)$ is the required $\gamma = \frac{\pu}{\sigma^2}$, such that the information sum-rate is $R$ in the absence of AOOBIs.}

\par {Similarly, let $\Rpcsit(\bm \Gamma_{\text{b}},R)$ be the information sum-rate achieved with $\gamma = \gamapcsi(R)$, in the presence of AOOBIs. Therefore, $\Rpcsit(\bm \Gamma_{\text{b}}, R) \Define \Rpcsi(\gamapcsi(R), \bm \Gamma_{\text{b}})$, $\forall \,\, \bm \Gamma_{\text{b}} > 0$ and $0 < R < \Rpcsi(\infty,\bm 0)$. Finally, for a given $0 < R^{\prime} < R$ and $0 < R < \Rpcsi(\infty,\bm 0)$, let $\gamabp(R, R^{\prime})$ be the maximum allowable total AOOBI power such that the information sum-rate in the presence of AOOBIs is $R^{\prime}$, i.e., $R^{\prime} = \Rpcsit(\frac{\gamabp(R, R^{\prime})}{\mathcal{I}}\bm 1, R) = \Rpcsi(\gamapcsi(R),\frac{\gamabp(R, R^{\prime})}{\mathcal{I}}\bm 1)$. For a given $(R, R^{\prime})$, we are interested in the maximum allowable ratio (MAR) $r_{\text{b}}$, i.e., from \eqref{eq:filtcut} we have}

\vspace{-0.6 cm}

\begin{IEEEeqnarray}{rCl}
\label{eq:pcsirb}
r_{\text{b, pcsi}} & = & {\gamabp(R, R^{\prime})} \Big/{\Big(1 + \gamapcsi(R) \sum_{q = 1}^{K} \beta_q \Big)} \,\,\,\, .
\IEEEeqnarraynumspace
\end{IEEEeqnarray}

\vspace{-0.35 cm}

\begin{proposition}
\label{pcsifiltcut}
For any given fixed $(R, R^{\prime})$, such that $0 < R^{\prime} < R < \Rpcsi(\infty, \bm 0)$, and fixed $K$, it can be shown that the ratio $r_{\text{b,pcsi}}$ in \eqref{eq:pcsirb} converges to a positive constant as $M \to \infty$.
\end{proposition}

\begin{IEEEproof}
We know that for any $0 < R < \Rpcsi(\infty,\bm 0)$, $R = \Rpcsi(\gamapcsi(R), \bm 0)$, i.e., from \eqref{eq:pcsisumrate} we get

\vspace{-0.5 cm}

{\begin{IEEEeqnarray}{rCl}
\label{eq:prop1}
R & = & \sum\limits_{k = 1}^{K} \log_2 \Bigg(1 + \frac{1}{\frac{1}{M}\big(1 + \sum_{q = 1, q\neq k}^{K}\frac{\beta_q}{\beta_k}\big) + \frac{1}{\beta_k M \, \gamapcsi(R)}}\Bigg).
\IEEEeqnarraynumspace
\end{IEEEeqnarray}}\normalsize

\vspace{-0.15 cm}

\indent Taking limit $M \to \infty$ on both sides of \eqref{eq:prop1} since $R$ is fixed, we have $R \, = \, \sum_{k = 1}^{K} \log_2 \left(1 + \beta_k \lim\limits_{M \to \infty} (M \, \gamapcsi(R))\right)$. This implies that $\lim\limits_{M \to \infty} M \gamapcsi(R) = c > 0$ (constant). Using this in \eqref{eq:pcsirb}, we obtain

\vspace{-0.7 cm}

\begin{IEEEeqnarray}{rCl}
\label{eq:pcsirblim}
\nonumber \lim\limits_{M \to \infty} r_{\text{b,pcsi}} & = & \lim\limits_{M \to \infty} \frac{\gamabp(R, R^{\prime})}{1 + \sum_{q = 1}^{K} \frac{\beta_q}{M} (\underbrace{M \gamapcsi(R)}_{ = \, c, \, \text{with}\, M \to \infty})}\\
& = &  \lim\limits_{M \to \infty} \gamabp(R, R^{\prime}) \, .
\IEEEeqnarraynumspace
\end{IEEEeqnarray}

\vspace{-0.2 cm}

\indent Next we analyze $\lim\limits_{M \to \infty} \gamabp(R, R^{\prime})$. We know that, $\gamabp(R, R^{\prime})$ satisfies $\Rpcsi(\gamapcsi(R),\frac{\gamabp(R, R^{\prime})}{\mathcal{I}} \bm 1) = R^{\prime}$ [see the paragraph before \eqref{eq:pcsirb}]. Using \eqref{eq:pcsisumrate}, we have

\vspace{-0.4 cm}

{\begin{IEEEeqnarray}{rCl}
\label{eq:prop2}
R^{\prime} = \sum\limits_{k = 1}^{K} \log_2 \Bigg(1 + \frac{1}{\frac{\big(1 + \sum_{q = 1, q\neq k}^{K}\frac{\beta_q}{\beta_k}\big)}{M } + \frac{1 + \gamabp(R,R^{\prime})}{\beta_k M \, \gamapcsi(R)}}\Bigg).
\IEEEeqnarraynumspace
\end{IEEEeqnarray}}\normalsize

\vspace{-0.15 cm}

\indent Taking limit as $M \to \infty$ on both sides of \eqref{eq:prop2}, we get

\vspace{-0.5 cm}

\begin{IEEEeqnarray}{rCl}
\label{eq:prop3}
R^{\prime} & = & \sum\limits_{k = 1}^{K} \log_2 \left(1 + \lim\limits_{M \to \infty} \frac{\beta_k c}{1 + \gamabp(R,R^{\prime})}\right),
\end{IEEEeqnarray}

\vspace{-0.2 cm}

\noindent since $R^{\prime}$ is fixed and $\lim\limits_{M \to \infty} \, M \gamapcsi(R) = c > 0$ (see the paragraph after \eqref{eq:prop1}). From \eqref{eq:prop3} it follows that for a given $(R,R^{\prime})$, $\lim\limits_{M \to \infty} \gamabp(R,R^{\prime}) = c^{\prime} > 0$ (constant). Using this in \eqref{eq:pcsirblim}, we have $\lim\limits_{M \to \infty} \, r_{\text{b,pcsi}} = c^{\prime} > 0$ (constant). \hfill \IEEEQEDhere
\end{IEEEproof}

\small{\begin{figure*}[!b]
\hrulefill
{{\begin{IEEEeqnarray}{lCl}
\label{eq:sinrkimcsi}
\text{SINR}_{k} (\gamma, \bm \Gamma_{\text{b}}) = {\left(\substack{\frac{1}{M}\Big[1 + \frac{2(\tau \gamma \beta_k \bm 1^T \bm \Gamma_{\text{b}}+\tau \gamma \beta_k + \bm 1^T \bm \Gamma_{\text{b}}) \, + \, (M + 1)\bm \Gamma_{\text{b}}^T\bm \Gamma_{\text{b}} }{(\tau \gamma \beta_k + \bm 1^T \bm \Gamma_{\text{b}} + 1)^2}\Big] \, + \, \Big(\frac{\tau \gamma \beta_k + \bm 1^T \bm \Gamma_{\text{b}} + 1}{M \tau \gamma^2 \beta_k^2}\Big)\bm 1^T \bm \Gamma_{\text{b}} \Big[1 + \frac{\frac{M}{\bm 1^T \bm \Gamma_{\text{b}}}\bm \Gamma_{\text{b}}^T\bm \Gamma_{\text{b}}}{\tau \gamma \beta_k + \bm 1^T \bm \Gamma_{\text{b}} + 1}\Big] \, + \, \frac{\tau \gamma \beta_k + \bm 1^T \bm \Gamma_{\text{b}} + 1}{M \tau \gamma^2 \beta_k^2}\\
 \,+ \, \sum\limits_{q = 1, q\neq k}^{K}\Big(\frac{\tau \gamma \beta_k + \bm 1^T \bm \Gamma_{\text{b}} + 1}{\tau \gamma \beta_q + \bm 1^T \bm \Gamma_{\text{b}} + 1}\Big)\frac{\beta_q^2}{M \beta_k^2} \Big[1 + \frac{M \bm \Gamma_{\text{b}}^T\bm \Gamma_{\text{b}}}{(\tau \gamma \beta_k + \bm 1^T \bm \Gamma_{\text{b}} +1)(\tau \gamma \beta_q + \bm 1^T \bm \Gamma_{\text{b}} +1)} \Big]\,+ \,\sum\limits_{q = 1}^{K} \Big(\frac{\tau \gamma \beta_k + \bm 1^T \bm \Gamma_{\text{b}} + 1}{\tau \gamma \beta_q + \bm 1^T \bm \Gamma_{\text{b}} + 1}\Big)\Big(\frac{\beta_q (\bm 1^T \bm \Gamma_{\text{b}} + 1)}{M \tau \gamma \beta_k^2}\Big)\,\, }\right)}^{-1}\\
 \label{eq:sinrkinf}
 \lim\limits_{M \to \infty} \, \text{SINR}_k \left(\gamacsi(R),\frac{\gamabi}{\mathcal{I}}\bm 1 \right)  =  {\left\{\lim\limits_{M \to \infty} \, \left(\substack{\frac{2\gamabi + (M+1)\frac{(\gamabi)^2}{\mathcal{I}}}{M(\gamabi + 1)^2} \, +  \, \sum\limits_{q = 1, q \neq k}^{K}\, \frac{\beta_q^2 \, \frac{(\gamabi)^2}{\mathcal{I}}}{\beta_k^2 \, (\gamabi + 1)^2} \, + \, \sum\limits_{q = 1}^{K} \, \frac{\beta_q \, (\gamabi + 1)}{\sqrt{M}\tau \, c \, \beta_k^2}\\ \, + \, \underbrace{\frac{1}{\tau \, c^2 \, \beta_k^2}M \frac{(\gamabi)^2}{\mathcal{I}}\,}_{\Define \, \text{T}_{\text{b}}} \, + \, \frac{(\gamabi + 1)^2}{\tau \, c^2 \, \beta_k^2}\,}\right)\right\}}^{-1}.
\IEEEeqnarraynumspace
\end{IEEEeqnarray}}}
\end{figure*}}\normalsize

\begin{remark}
\label{filtcutM}
\normalfont In the following, we explain the result in Proposition~\ref{pcsifiltcut}. In the absence of AOOBI signals, we know that for a fixed sum-rate $R$, the required $\gamma = \gamapcsi(R)$ decreases as $\frac{1}{M}$, with increasing $M \to \infty$ \cite{Ngo1}. Hence for sufficiently large $M$, in the absence of AOOBI signals, the total received in-band power (RIBP) at any BS antenna is dominated by AWGN, i.e., total RIBP = $\sigma^2 \big(1 + \gamma \sum_{q = 1}^{K} \beta_q\big) \approx \sigma^2$ (constant). 

\par In the presence of AOOBIs having a total aliased power $\sum_{i = 1}^{\mathcal{I}} p_i$, the information sum-rate decreases. For sufficiently large $M$, the total effective in-band noise power (EINP) therefore increases from $\sigma^2$ to $(\sigma^2 + \sum_{i = 1}^{\mathcal{I}} p_i) = \sigma^2 (1 + \gamab)$, where $\gamab = \sum_{i = 1}^{\mathcal{I}} p_i/\sigma^2 = \sum_{i = 1}^{\mathcal{I}} \gamma_i = \bm 1^T \bm \Gamma_{\text{b}}$ (note that $\gamma_i = p_i/\sigma^2$). With $M \to \infty$, we now have three possibilities: (a) if $\gamab \to 0$ with $M \to \infty$, then for sufficiently large $M$, the total EINP will be almost $\sigma^2$, i.e., almost same as that in the absence of AOOBIs. Therefore the information sum-rate in the presence of AOOBIs is the same as the information sum-rate in its absence; (b) if $\gamab$ increases unboundedly with $M \to \infty$, then EINP $\to \infty$ and therefore, the sum-rate in the presence of AOOBIs will converge to $0$; (c) if $\gamab$ converges to a positive constant as $M \to \infty$, then the EINP converges to a constant greater than $\sigma^2$, and therefore the information sum-rate in the presence of AOOBIs converges to a constant strictly less than the information sum-rate in its absence.

\par From the above discussion, it is clear that the information sum-rate in the presence of AOOBIs converges to a constant as $M \to \infty$, if and only if the total AOOBI power converges to a constant as $M \to \infty$. We know that the required BPF attenuation in the OOB region depends on the MAR of the total AOOBI power to the RIBP before aliasing, i.e., $r_{\text{b,pcsi}}$ (see \eqref{eq:pcsirb}). Since with $M \to \infty$, the RIBP before aliasing converges to $\sigma^2$ ($\because \gamma \propto 1/M$) and the total AOOBI power must converge to a constant, it follows that the required BPF attenuation converges to a constant as $M \to \infty$. \hfill \qed
\end{remark}

\vspace{-0.5 cm}

\section{Imperfect CSI Scenario}

For the imperfect CSI scenario, the first $\tau< N_u$ channel uses are used for pilot transmission from the UTs to the BS (training phase). BS uses these pilots for estimation of channel gain coefficients for different UTs. The UL data communication starts from $t = \tau$-th channel use. During pilot transmission, $s_k [t] = \sqrt{\tau \pu} \, \phi_k[t]$ ($t = 0, 1, \ldots, \tau - 1$) is the transmitted pilot signal from the $k^{\text{th}}$ UT. The pilot sequence transmitted by different UTs are orthogonal to each other. Let $\bm \phi_k \Define (\phi_k[0], \phi_k[1], \cdots, \phi_k[\tau - 1])^T \in \C^{\tau \times 1}$ be the vector of pilot symbols transmitted by the $k^{\text{th}}$ UT. Then $\bm \phi_k^H \, \bm \phi_j \, = \, 0$ if $k \neq j$ and $\bm \phi_k^H \, \bm \phi_j \, = \, 1$, if $k = j$, $\forall (k,j) \in \{1, 2, \ldots, K\}$. Now let $\bm \Phi \in \C^{\tau \times K}$ be defined such that $\bm \phi_k$ is its $k^{\text{th}}$ column. Then, $\bm \Phi^T  \bm \Phi^{\ast} = \bm I_K$. Moreover, during the UL data communication (i.e. after the transmission of pilots), we have $s_k[t] = \sqrt{\pu} \, x_k[t]$ ($t = \tau, \tau + 1, \ldots, N_u - 1$), where $x_k[t] \sim\mathcal{C}\mathcal{N}(0,1)$ is the i.i.d. information symbol transmitted from the $k^{\text{th}}$ UT at time $t$.

\vspace{-0.5 cm}

\subsection{LMMSE Channel Estimation}


Based on the pilot signals received during the training phase, the BS computes linear minimum mean square estimate (LMMSE) of the channel gain coefficients. Using \eqref{eq:rxsig}, the received pilot matrix for all $M$ antennas is given by $\bm R_p = \big[\, r_m[t] \, \big]_{M \times \tau} = \sqrt{\tau \pu}\, \bm H \bm \Phi^T + \sum_{i = 1}^{\mathcal{I}} \, \bm g_i \bm u_i^T + \bm W$, where $\bm W \Define \big[ \, w_m[t] \, \big]_{M \times \tau}$, $\bm H \Define \big[ \, h_{mk} \, \big]_{M \times K}$ and $\bm u_i \Define (u_i[0], u_i[1], \cdots, u_i[\tau - 1])^T$. Let $\widehat{h}_{mk}$ denote the LMMSE of the channel gain coefficient $h_{mk}$. Therefore, the LMMSE of the channel gain matrix $\bm H$ is given by $\widehat{\bm H} \Define \big[ \, \widehat{h}_{mk} \, \big]_{M \times K} = \bm R_p \bm \Phi^{\ast} \widetilde{\bm D} \, = \, (\sqrt{\tau \pu}\, \bm H + \sum_{i = 1}^{\mathcal{I}} \, \bm g_i \widetilde{\bm u}_i^T + \widetilde{\bm N}) \widetilde{\bm D}$, where $\widetilde{\bm u}_i \Define \bm \Phi^H \bm u_i$ and $\widetilde{\bm D} \Define \frac{1}{\sqrt{\tau \pu}} \Big(\bm I_K + \frac{1 + \, \bm 1^T \bm \Gamma_{\text{b}}}{\tau \gamma}\bm D^{-1}\Big)^{-1}$. Here $\bm D \Define \text{diag}(\beta_1, \beta_2, \cdots, \beta_K)$, $\widetilde{\bm N} \Define \bm W \bm \Phi^{\ast}$, $\gamma = \pu/\sigma^2$ and $\bm \Gamma_{\text{b}} = (\gamma_1, \gamma_2, \cdots, \gamma_{\mathcal{I}})^T$.


\subsection{MAR Analysis}

With MRC receiver processing at the BS, the detected information symbol from the $k^{\text{th}}$ UT at time $t$ is given by

\vspace{-0.55 cm}

\small{{\begin{IEEEeqnarray}{rCl}
\label{eq:detsigimcsi}
\nonumber \widehat{x}_k[t] & = & \,\,\,\,\,\, \sum\limits_{m = 1}^{M}\widehat{h}_{mk}^{\ast}\, r_m[t] = \sqrt{\pu}\, \widehat{\bm h}_k^{H} \sum\limits_{q = 1}^{K} \bm h_q \, x_q[t]\\
\nonumber & & \,\,\,\,\,\,\,\,\,\,\,\,\,\,\,\,\,\,\,\,\,\,+ \underbrace{\sum\limits_{i = 1}^{\mathcal{I}}\, \widehat{\bm h}_k^{H} \, \bm g_i \, u_i[t]}_{= \, \text{BL}_k[t]} + \underbrace{\widehat{\bm h}_k^{H}\, \bm w[t]}_{= \, \text{EN}_k[t]}\\
\nonumber & = & \underbrace{\sqrt{\pu} \E\Big[||\widehat{\bm h}_k||^2\Big]\, x_k[t]}_{= \, \text{ES}_k[t]} + \underbrace{\sqrt{\pu}\Big(||\widehat{\bm h}_k||^2 - \E\Big[||\widehat{\bm h}_k||^2\Big]\Big)x_k[t]}_{= \, \text{SIF}_k[t]} \\
\nonumber & & \,\,\,\,\,\,\,\,\, + \underbrace{\sqrt{\pu}\widehat{\bm h}_k^H \left(\sum\limits_{q = 1, q \neq k}^{K}(\widehat{\bm h}_q - \bm \epsilon_q) \, x_q[t] - \bm \epsilon_k x_k[t]\right)}_{= \, \text{MUI}_k[t]}\\
& & \,\,\,\,\,\,\,\,\,\,\,\,\,\,\,\,\,\,\,\,\,\,\,\,\,\,\,\,\,\,\,\,\,\,\,\,\,\,\,\,\,\,\,\, + \, \text{BL}_k[t]+ \text{EN}_k[t] \,\, ,
\IEEEeqnarraynumspace
\end{IEEEeqnarray}}}\normalsize

\vspace{-0.5 cm}

\noindent where $t \in \{\tau, \tau+1, \ldots, N_u - 1\}$, $\widehat{\bm h}_k \Define (\widehat{h}_{1k}, \widehat{h}_{2k}, \cdots, \widehat{h}_{Mk})^T$ and $\bm \epsilon_k \Define \widehat{\bm h}_k - \bm h_k \, \in \C^{M \times 1}$ is the LMMSE channel estimation error vector for the $k^{\text{th}}$ UT. Here $\text{EW}_k[t] = \text{SIF}_k[t] + \text{MUI}_k[t] + \text{BL}_k[t] + \text{EN}_k[t]$ is the overall noise. It can be shown that $\E[\text{EW}_k^{\ast}[t] \, x_k[t]] = 0, \, \forall \, k = 1, 2, \ldots,K$ and $t = \tau, \tau +1, \ldots, N_u - 1$, i.e., the effective noise $\text{EW}_k[t]$ is uncorrelated with the Gaussian information symbol $x_k[t]$. Since the worst case uncorrelated noise (in terms of mutual information) is Gaussian distributed with the same variance as that of $\text{EW}_k[t]$ \cite{Hasibi2}, a lower bound on the information rate would be $I(\widehat{x}_k[t];x_k[t]) \geq \log_2 \big(1 + \text{SINR}_k(\gamma, \bm \Gamma_{\text{b}})\big)$, where $\text{SINR}_k(\gamma, \bm \Gamma_{\text{b}}) \, \Define \, \E[\, |\text{ES}_k[t]|^2\,]/\E[\, |\text{EW}_k[t]|^2\,]$ is given by \eqref{eq:sinrkimcsi} at the bottom of the page.\footnote[5]{This coding strategy has also been used in some earlier works \cite{Hasibi2,Phasenoise}.} The sum-rate is then given by $\Rcsi(\gamma, \bm \Gamma_{\text{b}}) \Define \big(1 - \frac{\tau}{N_u}\big)\, \sum_{k = 1}^{K}\, \log_2 \big(1 + \text{SINR}_k(\gamma, \bm \Gamma_{\text{b}})\big)$.

\par {Let $\Rcsi(\infty,\bm 0) \Define \lim\limits_{\gamma \to \infty} \, \Rcsi(\gamma, \bm \Gamma_{\text{b}} = \bm 0)$. In the absence of aliased OOBIs (AOOBIs), for any $0 < R < \Rcsi(\infty, \bm 0)$, we can define $\gamma = \gamacsi(R)$ to be the unique solution to $\Rcsi(\gamma, \bm 0) = R$, i.e., $\gamacsi(R)$ is the required transmit SNR $\gamma = \frac{\pu}{\sigma^2}$ to achieve a information sum-rate $R$. Also let $\widetilde{R}_{\text{icsi}}(\bm \Gamma_{\text{b}}, R) \Define \Rcsi(\gamacsi(R), \bm \Gamma_{\text{b}})$ be the information sum-rate achieved with $\gamma = \gamacsi(R)$ in the presence of AOOBIs. From \eqref{eq:filtcut}, it is clear that the maximum allowable ratio (MAR) $r_{\text{b}}$ depends on $\bm \Gamma_{\text{b}}$ only through $\gamab = \bm 1^T \bm \Gamma_{\text{b}}$. Therefore to compute $r_{\text{b}}$, in the following, for a given information sum-rate $R$ in the absence of AOOBIs, and sum-rate $R^{\prime} < R$ in the presence of AOOBIs, we maximize $\bm 1^T \bm \Gamma_{\text{b}}$ over all vectors $\bm \Gamma_{\text{b}}$ such that $R^{\prime} = \widetilde{R}_{\text{icsi}}(\bm \Gamma_{\text{b}}, R) = \Rcsi(\gamacsi(R), \bm \Gamma_{\text{b}})$. It turns out that the maximizing vector $\bm \Gamma_{\text{b}}^{\ast}(R, R^{\prime}) \Define (\gamma_1^{\ast}, \gamma_2^{\ast}, \cdots, \gamma_{\mathcal{I}}^{\ast})^T$ is such that $\gamma_1^{\ast} = \cdots = \gamma_{\mathcal{I}}^{\ast}$.\footnote[6]{{From the SINR expression in \eqref{eq:sinrkimcsi} it follows that $\forall \, \bm \Gamma_{\text{b}}$ vectors having the same value of $\bm 1^T \bm \Gamma_{\text{b}}$, the highest SINR is achieved when all components of $\bm \Gamma_{\text{b}}$ are equal}.} For the given $(R, R^{\prime})$, let $\gamma_b(R, R^{\prime}) \Define \bm 1^T \bm \Gamma_{\text{b}}^{\ast} (R, R^{\prime})$. Note that $\bm \Gamma_{\text{b}}^{\ast}(R, R^{\prime}) = \frac{\gamma_b(R, R^{\prime})}{\mathcal{I}} \bm 1$. Therefore for a desired $(R, R^{\prime})$, the MAR $r_{\text{b}}$ [see \eqref{eq:filtcut}] in the imperfect CSI scenario is given by}

\vspace{-0.55 cm}

\begin{IEEEeqnarray}{rCl}
\label{eq:imcsirb}
r_{\text{b}} & = & {\gamab(R,R^{\prime})} \Big /{\Big (1 + \gamacsi(R) \sum_{q = 1}^{K} \beta_q \Big )}.
\IEEEeqnarraynumspace
\end{IEEEeqnarray}

\vspace{-0.2 cm}

\indent In the following theorem, we present an interesting result on the variation of $r_{\text{b}}$ with increasing $M \to \infty$.


\begin{theorem}
\label{imcsifiltcut}
For any given $(R, R^{\prime})$, such that $0 < R^{\prime} < R < \Rcsi(\infty,\bm 0)$, and fixed $K$, the ratio $r_{\text{b}}$, defined in \eqref{eq:imcsirb} decreases as $\frac{1}{\sqrt{M}}$, as $M \to \infty$, i.e., $\lim\limits_{M \to \infty} \, \sqrt{M} \, r_{\text{b}} = $ constant.
\end{theorem}

\begin{IEEEproof}
See Appendix~A. \hfill \IEEEQEDhere
\end{IEEEproof}

\begin{figure}[!t]
\vspace{-0.85 cm}
\centering
\includegraphics[width= 3.4 in, height= 1.95 in]{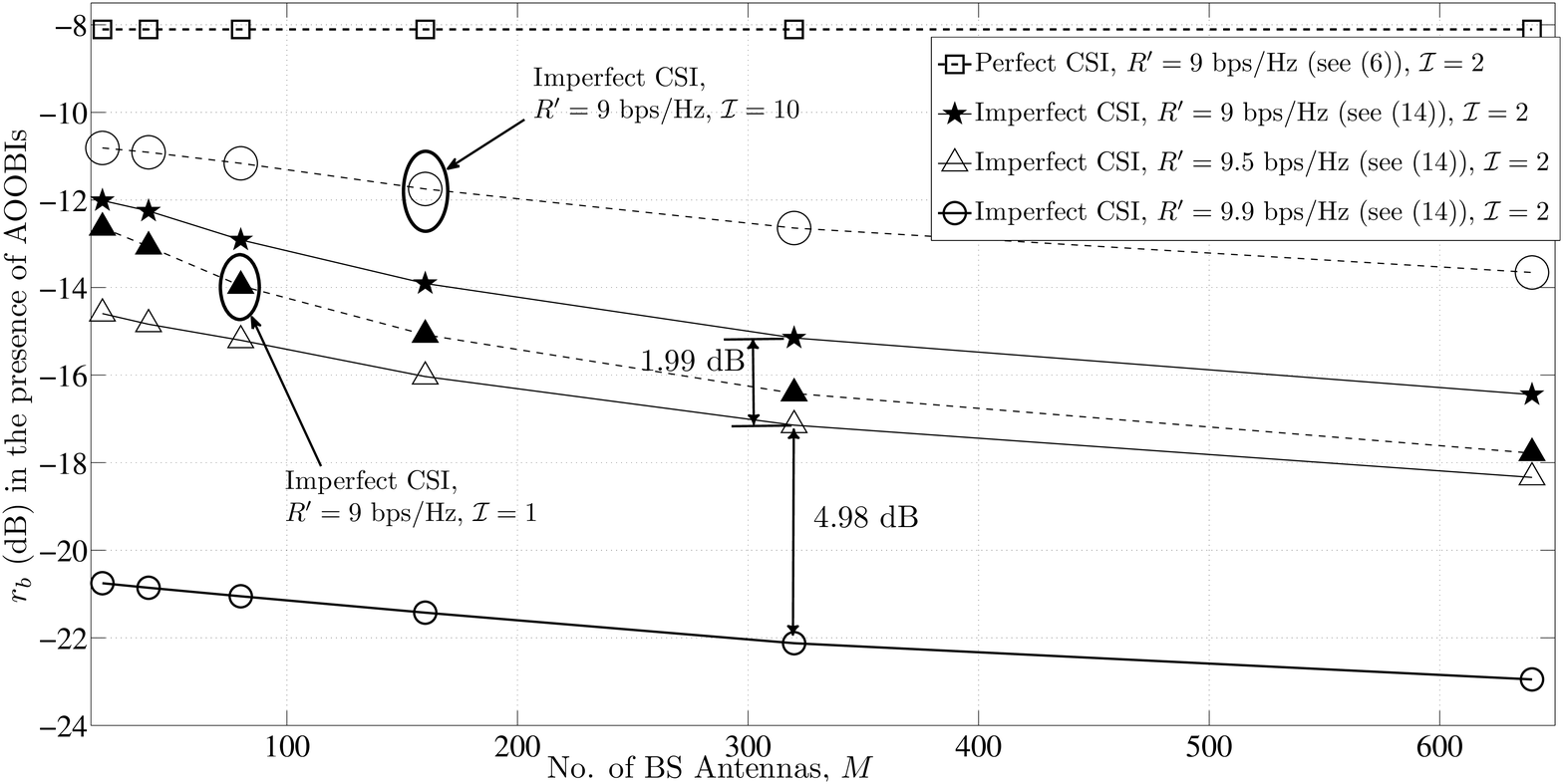}
\caption {{Plot of $r_{\text{b}}$ versus $M$ for fixed $K = 10$ and $N_u = 100$. Fixed desired sum-rate of $R = 10$ bpcu in the absence of AOOBIs and a fixed desired sum-rate of $R^{\prime}$ in the presence of AOOBIs.}}
\label{fig:fixlossvarM}
\vspace{-0.7 cm}
\end{figure}

\begin{remark}
\label{remimcsifiltcut}
\normalfont In the following we explain the result in Theorem~\ref{imcsifiltcut}. In the imperfect CSI scenario, the channel estimates acquired at the BS are corrupted by the AOOBI signals received along with the uplink pilots. Assuming the channel gain from the AOOBI to the BS to be same during both pilot transmission and UL data communication, {maximum ratio diversity combining (i.e. MRC)} at the BS, also leads to combining of the channel gains from the AOOBIs to the BS (note the term $\text{BL}_k[t]$ in the second line of \eqref{eq:detsigimcsi}). Due to this combining, the post-combining total AOOBI power increases with increasing $M$ ($\propto M\sum_{i = 1}^{\mathcal{I}} \gamma_i^2$). Hence to achieve a fixed sum-rate, we must decrease the power of each AOOBI, i.e., the total AOOBI power $\gamab = \sum_{i = 1}^{\mathcal{I}} \gamma_i$ must decrease with increasing $M$ in order that the effective sum-rate is almost constant for sufficiently large $M$. From our analysis, we see that $\gamab$ must be decreased as $\frac{1}{\sqrt{M}}$ with $M \to \infty$. Since with $M \to \infty$ the total received in-band power before aliasing (RIBP) is almost constant, i.e., $\big(\sigma^2 + \pu \sum_{q = 1}^{K} \, \beta_q \big) \, \stackrel{\text{\tiny{$M \to \infty$}}}{\longrightarrow} \, \sigma^2$ ($\because \gamma = \frac{\pu}{\sigma^2} \propto \frac{1}{\sqrt{M}}$), it follows that the MAR $r_{\text{b}}$ must decrease as $\frac{1}{\sqrt{M}}$ with $M \to \infty$ (see also Fig.~\ref{fig:fixlossvarM}). \hfill \qed
\end{remark}

Note that this is contrary to the result obtained with perfect CSI, where with $M \to \infty$, the MAR $r_{\text{b}}$ converges to a constant. The result in the imperfect CSI scenario (Theorem~\ref{imcsifiltcut}) implies that the required BPF attenuation in the OOB region must increase as $\mathcal{O}(\sqrt{M})$ with increasing $M$, which in turn would increase the design complexity and hardware cost. This imposes a practical limit on how large $M$ can be, depending on the trade-off between the required BPF attenuation, channel bandwidth, hardware cost and power consumption.

\vspace{-0.8 cm}

\section{Numerical Results and Discussions}
In this section we numerically study variation of the MAR $r_{\text{b}}$ with increasing number of BS antennas. The analytical expression of $r_{\text{b}}$ is given by \eqref{eq:imcsirb} for the imperfect CSI scenario and by \eqref{eq:pcsirb} for the perfect CSI scenario respectively. For the imperfect CSI scenario, we perform an additional optimization with respect to the training duration $\tau$.{\footnote[7]{{We know that for a given $(R, R^{\prime})$, the MAR $r_{\text{b}}$ depends on $\gamab(R, R^{\prime})$ which is the maximum allowable total AOOBI power such that the sum-rate in the presence of AOOBIs is $R^{\prime}$ and it is $R$ in their absence. Since the sum-rates $(R, R^{\prime})$ depend on $\tau$, we therefore numerically compute $\Rcsi(\gamma, \bm \Gamma_{\text{b}})$ as $\Rcsi(\gamma,\bm \Gamma_{\text{b}}) = \max\limits_{K \leq \tau \leq N_u-1} \big(1 - \frac{\tau}{N_u}\big) \, \sum_{k = 1}^{K} \, \log_2\big(1 + \text{SINR}_k(\gamma,\bm \Gamma_{\text{b}})\big)$, where $\text{SINR}_k(\gamma,\bm \Gamma_{\text{b}})$ is given by \eqref{eq:sinrkimcsi}.}}} We use the following system parameters: the number of UTs $K = 10$ and communication bandwidth $B_c = 200$ KHz. The channel coherence time is $T_c = 1$ ms and therefore duration of the coherence interval is $N_c = T_c\, B_c = 200$ channel uses. The duration of UL slot is $N_u = 100$ channel uses. The information sum-rate in the absence of AOOBIs is $R = 10$ bits per channel use (bpcu). Moreover for simplicity we assume $\beta_k = 1$, $\forall \, k = 1,2, \ldots, K$ and the number of AOOBIs $\mathcal{I} = 2$.

\par In Fig.~\ref{fig:fixlossvarM} we study the variation in $r_{\text{b}}$ with increasing number of BS antennas, $M$ for a fixed information sum-rate $R^{\prime} = 9$ bpcu in the presence of AOOBIs (i.e. $10\%$ fractional loss). In the perfect CSI scenario, we observe that $r_{\text{b}}$ converges to a constant with increasing $M \to \infty$ (see Proposition~\ref{pcsifiltcut}). Note that with imperfect CSI, with increasing $M \to \infty$, the decrease in $r_{\text{b}}$ is almost $1.5$ dB with every doubling in $M$ (see the curve with filled stars for $M = 160$ and $M = 320$). This supports the $\mathcal{O}(\sqrt{M})$ decrease, suggested in Theorem~\ref{imcsifiltcut}. We also plot $r_{\text{b}}$ for $R^{\prime} = 9.5$ and $9.9$ bpcu (i.e. $5\%$ and $1\%$ fractional loss). Note that for $M = 320$, by increasing the acceptable fractional loss from $1\%$ to $5\%$, we can relax the required BPF attenuation by a phenomenal $4.98$ dB. Further increase in acceptable fractional loss however gives only small relaxation in the required BPF attenuation. We also plot $r_{\text{b}}$ versus $M$ for $R^{\prime} = 9$ bpcu with $\mathcal{I} = 1, 10$ respectively. It is observed that the $\mathcal{O}(\sqrt{M})$ decrease holds true irrespective of the number of AOOBIs.

%






\vspace{-0.6 cm}
\appendix[Proof of Theorem~\ref{imcsifiltcut}]


From earlier works \cite{Ngo1}, it can be easily shown that with $M$ sufficiently large, i.e., $M \to \infty$, for any given $0 < R < \Rcsi(\infty, \bm 0)$, we have $\lim\limits_{M \to \infty} \, \sqrt{M} \gamacsi(R) = c > 0$ (constant), i.e., $\gamacsi(R)$ decreases as $1/\sqrt{M}$ with increasing $M$. Taking limit on both sides of \eqref{eq:imcsirb}, we get

\vspace{-0.55 cm}

\begin{IEEEeqnarray}{rCl}
\label{eq:imcsirbm}
\nonumber \lim\limits_{M \to \infty} \sqrt{M} \, r_{\text{b}} & = & \lim\limits_{M \to \infty} \frac{\sqrt{M} \, \gamabi}{1 + \underbrace{\sqrt{M}\, \gamacsi(R)}_{= \, c, \, \text{with} \, M \to \infty} \, \sum_{q = 1}^{K} \, \frac{\beta_q}{\sqrt{M}}}\\
& = & \lim\limits_{M \to \infty} \, \sqrt{M}\, \gamabi.
\end{IEEEeqnarray}

\vspace{-0.2 cm}

\indent Next we analyze $\lim\limits_{M \to \infty} \, \sqrt{M}\, \gamabi$. We know that for $0 < R^{\prime} < R < \Rcsi(\infty, \bm 0)$, $R^{\prime} = \Rcsi(\gamacsi(R), \frac{\gamabi}{\mathcal{I}}\bm 1) = \big(1 - {\tau}/{N_u}\big) \sum_{k = 1}^{K} \, \log_2 \big[1 + \text{SINR}_k(\gamacsi(R),\frac{\gamabi}{\mathcal{I}}\bm 1)\big]$ (see the paragraph before \eqref{eq:imcsirb}). Taking limit $M \to \infty$ in this expression and using $\lim\limits_{M \to \infty} \, \sqrt{M} \gamacsi(R) = c$, we have $R^{\prime} = \big(1 - \frac{\tau}{N_u}\big) \, \sum_{k = 1}^{K} \, \log_2 \big(1 + \lim\limits_{M \to \infty} \,\text{SINR}_k(\gamacsi(R),\frac{\gamabi}{\mathcal{I}}\bm 1)\big)$, where $\lim\limits_{M \to \infty} \,\text{SINR}_k(\gamacsi(R), \frac{\gamabi}{\mathcal{I}}\bm 1)$ is given by \eqref{eq:sinrkinf}, at the bottom of the previous page. Since $R^{\prime} < R$ is fixed, we have $\lim\limits_{M \to \infty} \,\text{SINR}_k(\gamacsi(R),\frac{\gamabi}{\mathcal{I}}\bm 1)=$ constant $>0$.

\par We next show that $\lim\limits_{M \to \infty} \text{SINR}_k (\gamacsi(R), \frac{\gamabi}{\mathcal{I}}\bm 1) = \text{constant} > 0 \Longleftrightarrow$  $\lim\limits_{M \to \infty} \sqrt{M} \gamabi = \text{constant}> 0$. If $\lim\limits_{M \to \infty} \sqrt{M} \gamabi = c_{\text{b}} >0$ (constant), then from \eqref{eq:sinrkinf} we have $\lim\limits_{M \to \infty} \, \text{SINR}_k (\gamacsi(R), \frac{\gamabi}{\mathcal{I}}\bm 1) = \frac{\tau \, c^2 \, \beta_k^2}{1 + c_{\text{b}}^2/\mathcal{I}} > 0$ (constant). For the reverse statement, it suffices to show that if $\lim_{M \to \infty} \sqrt{M} \gamab(R, R^{\prime}) = \infty$, then the SINR converges to zero as $M \to \infty$. Towards this end, we see that if $\lim_{M \to \infty} \sqrt{M} \gamab(R, R^{\prime}) = \infty$, then $\lim_{M \to \infty} T_{\text{b}} = \infty$ (where the term $T_{\text{b}}$ is defined in \eqref{eq:sinrkinf}). From this it follows that $\lim_{M \to \infty} \text{SINR}_k (\gamacsi(R), \frac{\gamab(R, R^{\prime})}{\mathcal{I}} \bm 1) = 0$. Finally, using the fact that $\lim\limits_{M \to \infty}\, \sqrt{M} \, \gamabi = $ constant $>0$ in \eqref{eq:imcsirbm}, we get $\lim\limits_{M \to \infty} \, \sqrt{M} \, r_{\text{b}} = $ constant $>0$.



%

%


\ifCLASSOPTIONcaptionsoff
  \newpage
\fi



%

\vspace{-0.35 cm}

\bibliographystyle{IEEEtran}
\bibliography{IEEEabrvn,mybibn}


\end{document}